\documentclass[aip,jcp,
			   reprint,
			   graphicx,
			   amsmath,amssymb
			   ]{revtex4-1}
\usepackage{graphicx}
\usepackage{cmap}
\usepackage[version=3]{mhchem}
\usepackage{sistyle}
\usepackage{xspace}
\usepackage[pdftex,colorlinks,citecolor={blue},linkcolor={red}, urlcolor={violet}]{hyperref}
\usepackage{times}
\usepackage{mathptmx}
\usepackage{geometry}
\usepackage{afterpage}

\usepackage[dvipsnames]{xcolor}

\begin{document}

\newcommand\Ih{I$_\text{h}$\xspace}
\newcommand\ie{\emph{i.e.}\xspace}

\title{Heterogeneous ice nucleation on silver-iodide-like surfaces}
\newcommand\PTCL{Physical and Theoretical Chemistry Laboratory, Department of Chemistry, University of Oxford, Oxford OX1 3QZ, United Kingdom}

\author{Guillaume Fraux}
\affiliation{\PTCL}
\affiliation{\'Ecole Normale Sup\'erieure de Paris, 45 rue d'Ulm, 75005 Paris, France}

\author{Jonathan P.K. Doye}
\affiliation{\PTCL}

\date{\today}

\maketitle

Ice nucleation is an important phenomenon influencing many aspects of our
environment, from climate \cite{Murray2012} to biological systems.
Indeed, understanding ``how ice forms'' has recently been identified as one of
the top ten open questions in ice science.\cite{Atkinson2013} Almost all ice
nucleation occurs heterogeneously, the process being initiated by an external
nucleation agent like a crystalline dust particle or organic matter. However,
we have relatively little knowledge about the microscopic mechanisms of this
transformation, because the infrequent but rapid nature of the ice nucleation
events and the small size of critical nuclei make these details hard to probe
experimentally.
Indeed, even for crystals that are good ice nucleants, little is known about
which surfaces, steps or defects may be the active nucleation sites.

Computer simulations can potentially help by providing insights into the
microscopic mechanisms. 
However, ice nucleation is hard to
simulate with all-atom models of water, as the natural dynamics of ice growth is
very slow at temperatures where the critical nucleus is of a size that is
reasonable to simulate.\cite{Reinhardt2013c,*Sanz2013b} Nevertheless,
heterogeneous ice nucleation has recently become increasingly probed by
computer
simulations,\cite{Yan2011,*Yan2012,*Yan2013,Cox2013,Lupi2014,*Lupi2014b,*Singh2014,Zhang2013d,*Zhang2014,Reinhardt2014b}
for example on kaolinite\cite{Cox2013} and
graphitic\cite{Lupi2014,*Lupi2014b,*Singh2014} surfaces.

Silver iodide (AgI) is an excellent ice nucleant that is widely used in cloud
seeding.  Its nucleation ability was discovered in 1947 by
Vonnegut\cite{Vonnegut1947} 
based on the structural similarity between $\beta$-AgI and ice
\Ih---both are based on the hexagonal diamond lattice and have a close lattice
match ($\sim$3\%).  In this paper, we consider ice nucleation at model AgI-like
surfaces with the aim of improving our fundamental understanding of how ionic
crystal surfaces can potentially initiate ice nucleation.  
Previous modelling
of the AgI-water system has focussed on the preferred water adsorption
sites,\cite{Fukuta1973,Hale1980,*Hale1980b} the structure of surface water
clusters\cite{Ward1982,*Ward1983} and layers,\cite{Taylor1993} and nucleation
of liquid water on AgI particles.\cite{Shevkunov2005} 

To model this system we use the TIP4P/2005\cite{Abascal2005} potential for
water, the Parrinello, Rahman, and Vashishta (PRV) potential for
AgI,\cite{Parrinello1983} and a potential with a Lennard-Jones term and a
Coulombic term for the AgI-water interactions where the parameters were taken
from Ref.~\onlinecite{Hale1980,*Hale1980b}.  However with this setup, we found
that, in contrast to real AgI, the crystal began to quickly dissolve in water,
and therefore, we constrained the \ce{Ag+} and \ce{I-} ions with a
harmonic potential to their equilibrium positions in a perfect AgI crystal,
with the spring constant set to $k = \SI{1000}{kcal\,mol^{-1}\,\AA^{-2}}$ 
instead of using the PRV potential.

Simulations were carried out $T = \SI{242}{K}$ which represents a supercooling
of approximately
$\SI{10}{K}$ for TIP4P/2005.  An AgI crystal slab of the correct orientation
was placed in the middle of the box, so as to expose the basal, prismatic or
normal faces, with water on both sides. We ran molecular dynamics simulations
at constant volume and temperature with repulsive walls at the top and the
bottom of the box and periodic boundary conditions in $x$ and $y$, with the
position of the walls chosen so that there was $\sim$\SI{20}{\AA} of empty
space at both ends of the box. 
We used a particle-particle particle-mesh (PPPM) solver -- modified to account
for the non-periodic condition in $z$ -- to deal with the long range Coulombic
interactions; all other potentials were cut off after \SI{8.5}{\AA}.

\begin{figure*}[t]
	\includegraphics[width=\textwidth]{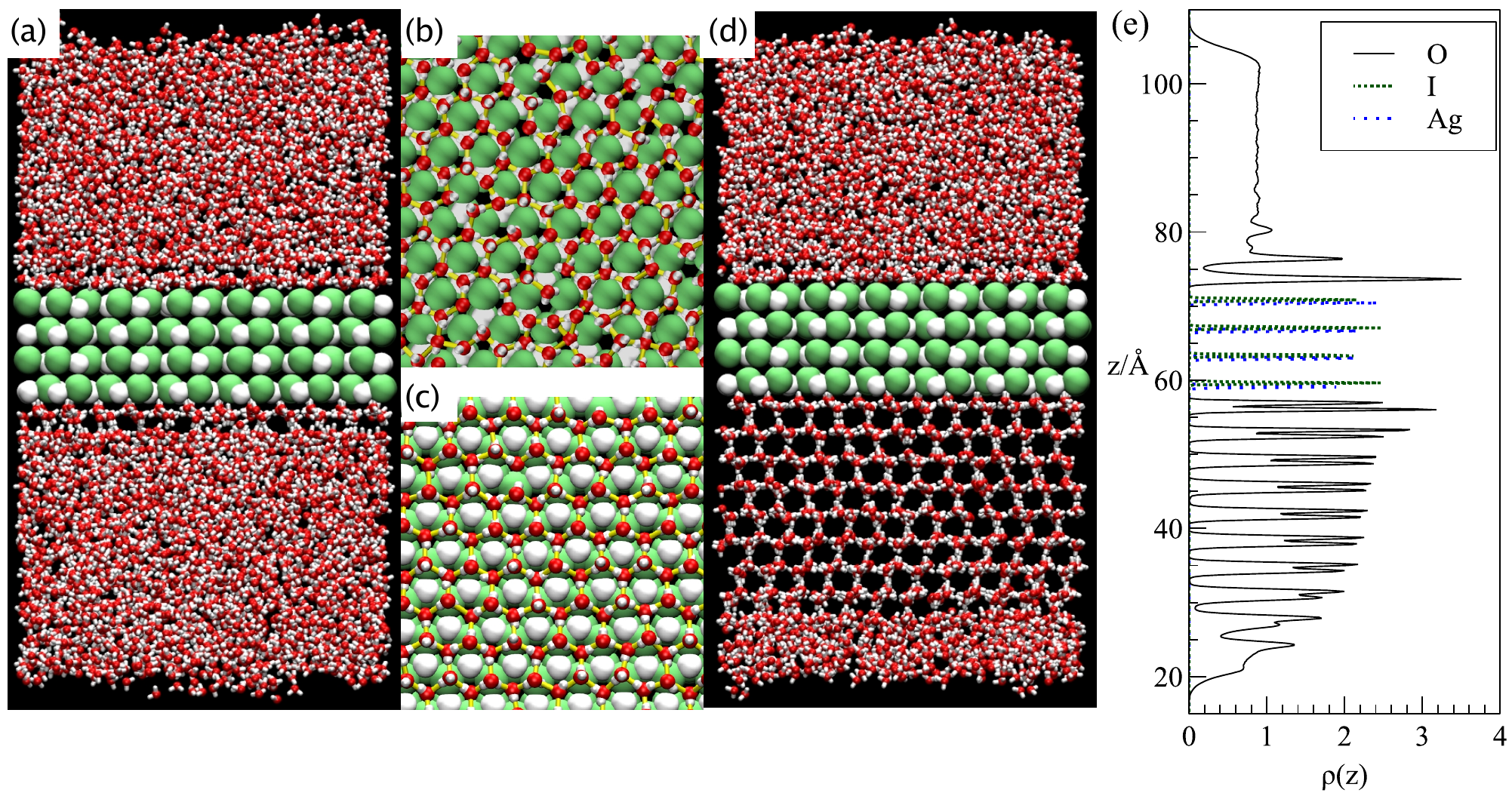}
	\caption{Ice nucleation on the basal faces of AgI. Snapshots after $t=\SI{3}{ps}$ of (a) the whole simulation box, and the adsorbed layers on the (b) \ce{I-}-terminated and (c) \ce{Ag+}-terminated faces. (d) Snapshot of the whole box and (e) local density variation along $z$ at $t=\SI{160}{ns}$. 
The box contains 4914 water molecules and 960 ions in the AgI crystal.}
	\label{fig}
\end{figure*}

We found that relatively facile ice nucleation and growth occured on the
\ce{Ag+}-terminated basal plane, but no ice formation was seen for any of the
other faces on our simulation time scales. The ice formation occurs via a
two-step mechanism. First, there is the fast formation of a hexagonal layer of
strongly adsorbed water molecules on the surface.\cite{Shevkunov2005} This
initial ice-like bilayer, illustrated in Figure~\ref{fig}(c), forms in less
than \SI{3}{ps}. Then, after a certain lag time (of the order of \SI{40}{ns}),
a nucleation event occurs and ice starts to grow on the top of this layer
(typically 1 layer per 10--$\SI{15}{ns}$).  After sufficient time, as
illustrated in Fig.\ \ref{fig}(d), most of the water above the
\ce{Ag+}-teminated face forms ice, and the corresponding density profile
(Fig~\ref{fig}(e)) shows a series of clear double peaks only at the
silver-terminated side, their split structure reflecting the ``chair''
conformations of the hexagons in these bilayers.

Interestingly, although a somewhat similar adsorbed layer formed on the
\ce{I-}-terminated side (see Fig.~\ref{fig}(b)),\cite{Taylor1993} this layer
was unable to initiate ice growth. On the \ce{Ag+}-terminated basal face, the
strong interaction between the \ce{Ag+} ions and the oxygens strongly constrains
the positions of the water molecules in the adsorbed layer, whereas the 
\ce{I-}-hydrogen interaction provides a weaker constraint on the adsorbed water
on the \ce{I-} face .  The greater disorder and flatter, less bilayer-like
character of the layer is evident from Fig.~1(a) and (b) and underlies its
inability to template ice growth.  Similarly, although water molecules strongly
adsorbed on the other faces (prism and normal) that we considered, these layers
are less obviously ice-like and are unable to act as templates for ice growth
on our simulation time scales.

The current report provides one of the first systems where heterogeneous ice
nucleation has been observed for an all-atom model of water in the absence of
an imposed electric field and shows how an ionic surface can efficiently
initiate ice nucleation \emph{via} a two-step mechanism. The surface must be
capable of templating the formation of a sufficiently ordered ice-like layer,
which will template further ice growth. A very similar mechanism has previously
been reported in Ref.~\onlinecite{Reinhardt2014b} for a monatomic water model
on model surfaces.

Although our results provide interesting fundamental insights into a
potential mechanism of heterogeneous nucleation on ionic surfaces, there are
two important caveats concerning its application to real silver iodide. Firstly,
the spring constant used to constrain the position of the ions in the crystal
is unrealistically large. For example, a value $k =
\SI{70}{kcal\,mol^{-1}\,\AA^{-2}}$ is needed to reproduce the width of the
peaks in the radial distribution function obtained with the PRV potential.
However, nucleation was never seen on the time scale (up to $\SI{260}{ns}$) of
our simulations with smaller $k$. Interestingly, this indicates that restricted
thermal motion is important for templating a sufficiently ordered layer that
will allow further ice growth. We note that most previous simulations of
heterogeneous ice nucleation have typically used a fixed substrate with no
thermal
motion.\cite{Lupi2014,*Lupi2014b,*Singh2014,Zhang2013d,*Zhang2014,Reinhardt2014b} 

Secondly, the basal plane of silver iodide has a non-zero surface dipole, and
it is well established that such surfaces are unstable in vacuum and should
reconstruct.\cite{Tasker1979} Indeed, when we simulate such an AgI slab in
vacuum using the PRV potential, it undergoes a surface and bulk reconstruction.
Hence, there is considerable uncertainty concerning the structure of AgI
basal planes in water. To probe this, a more realistic
description of the water-AgI system is needed.  

We acknowledge the use of Advanced Research Computing (ARC) in Oxford and 
helpful discussions with Flavio Romano, Aleks Reinhardt and Mark Wilson.\\


%

\end{document}